\documentclass[aps,11pt,nofootinbib,preprintnumbers,groupedaddress,superscriptaddress]{revtex4-1}

\usepackage{amssymb, amsfonts, amsmath, bm}
\usepackage[
]{graphicx}
\usepackage{color}

\definecolor{DarkGreen}{rgb}{0,0.3,0}

\definecolor{cobalt}{rgb}{0.0, 0.28, 0.67}

\usepackage[
]{hyperref}
\hypersetup{%
 setpagesize=false,
 bookmarksnumbered=true,%
 bookmarksopen=true,%
 colorlinks=true,%
 linkcolor=blue,%
 citecolor=red}

\begin{document}
\title{Chaos in a generalized Euler's three-body problem}
\author{Takahisa Igata}
\email{igata@post.kek.jp}
\affiliation{KEK Theory Center, 
Institute of Particle and Nuclear Studies, 
High Energy Accelerator Research Organization, Tsukuba 305-0801, Japan}
\date{\today}
\preprint{KEK-Cosmo-0272, KEK-TH-2302}

\begin{abstract}
Euler's three-body problem is the problem of solving for the motion of 
a particle moving in a Newtonian potential generated by two point sources fixed in space. 
This system is integrable in the Liouville sense.
We consider the Euler problem with the inverse-square potential, 
which can be seen as a natural generalization of the three-body problem to 
higher-dimensional Newtonian theory. 
We identify a family of stable stationary orbits in the generalized Euler problem. 
These orbits guarantee the existence of stable bound orbits. 
Applying the Poincar\'e map method to these orbits, 
we show that stable bound chaotic orbits appear.
As a result, we conclude that the generalized Euler problem is nonintegrable.
\end{abstract}
\maketitle

\section{Introduction}
\label{sec:1}
The Kepler problem in celestial mechanics---two bodies interact with each other 
by the inverse-square force law---has provided us 
with many insights into the motion of astrophysical bodies.
This problem was generalized by Euler 
to a certain three-body problem---motion of a particle subjected to 
the inverse-square force sourced by two fixed masses 
in space---which is known as Euler's three-body problem~\cite{Euler:1760}.

These problems have a common feature of the Liouville integrability.
The Kepler problem is integrable, i.e., 
there exist three constants of motion, energy, 
angular momentum around an axis, 
and the squared total angular momentum, 
commutable with each other by the Poisson brackets.
As a result, the equations of motion reduce to 
decoupled ordinary differential equations for each variables. 
In addition, there is a vector-type constant of motion in this system, 
the so-called Laplace-Runge-Lenz vector (see, e.g., Ref.~\cite{Goldstein:1980}). 
Since all six of those constants are Poisson commutable, 
the Kepler problem is superintegrable and is solved in an algebraic way. 
On the other hand, the Euler problem admits 
two constants of motion, energy and angular momentum.
Additionally, there is a nontrivial constant of 
a second-order polynomial in momentum, 
the so-called Whittaker constant~\cite{Whittaker:1917,Biscani:2016}.
These three constants are Poisson commutable with each other, 
and therefore, the system is integrable. 
The nontrivial constants in both cases are 
relevant to hidden symmetry developed in 
general relativity---the Killing tensors (see, e.g., Ref.~\cite{Yasui:2011pr}). 
Though the constants are composed by reducible Killing tensors 
(i.e., the solution to the Killing tensor equations in the Euclidean space is given as a linear combination of the symmetric tensor products of the Killing vectors.)
because of the maximal symmetry in the Euclidean space,
they still play an important role in terms of the integrability of particle dynamics.

Such nontrivial constants were discovered by exploring 
the separation of variables of the corresponding Hamilton-Jacobi equation. 
This insight was applied to the geodesic systems in the Kerr spacetime, 
and it was found that the separation of variables of 
the Hamilton-Jacobi equation occurs and 
there exists the Carter constant~\cite{Carter:1968rr}, 
which is relevant to the irreducible and nontrivial Killing tensor~\cite{Walker:1970un}. 
This discovery leads us to study particle motion in the Kerr geometry 
further in analytical ways~(e.g., Ref.~\cite{Kraniotis:2004cz}).

It is a natural but nontrivial question to ask whether integrability is 
preserved if the system is generalized somehow.
In generalizing an integrable Newtonian particle system to 
a general relativistic particle system, we often encounter examples 
where integrability breaks down.
For example, the dynamics of a massive particle
in the four-dimensional (4D) Majumdar-Papapetrou 
dihole spacetime~\cite{Majumdar:1947eu,Papaetrou:1947ib}---a generalization of 
the Euler problem to general relativity---is nonintegrable because it exhibits chaotic behavior~\cite{Contopoulos:1990,Contopoulos:1991,Yurtserver:1995,Dettmann:1994dj,Dettmann:1995ex,Alonso:2007ts,Shipley:2016omi}. 
In other relativistic Euler's three-body problems, 
chaotic orbits also appear in general~\cite{Cornish:1996de,deMoura:1999zd,Coelho:2009gy}.

Recently, with the progress in the study of 
higher-dimensional black holes~\cite{Emparan:2008eg}, 
the relationship between the hidden symmetry and the Liouville integrability 
has been gradually revealed through the dynamics of particles in higher dimensions~\cite{Frolov:2017kze}.
One of the generalizations of the Euler problem to 
higher-dimensional space was also discussed in the context of Newtonian theory, 
in which the spatial dimension was parameterized 
while the potential still
being inversely proportional to distance~\cite{Coulson:1967}.
In this case, even if the spatial dimension is arbitrary, 
there exists the same type of the Whittaker constant; 
in other words, the integrability is preserved. 
However, in generalization of Newtonian gravity 
to higher-dimensional space, the power-law of gravitational force 
should be modified accordingly.%
\footnote{The full three-body problem in Newtonian gravity 
with an inverse-square potential in $\mathbb{E}^4$ was 
discussed in Ref.~\cite{Jimenez-Lara:2003}.}
Therefore, in this paper, we consider the Euler problem with the inverse-square potential, 
which can be seen as a natural generalization of the three-body problem to 
higher-dimensional Newtonian theory.
The purpose of this paper is to clarify 
whether the integrable structure of the original Euler problem can be preserved 
even in such a generalized Euler's three-body problem.

It should be noted that the timelike geodesics 
in the higher-dimensional Majumdar-Papapetrou dihole spacetime show 
chaotic behavior~\cite{Hanan:2006uf}.
Thus, we may also ask whether the Newtonian limit of this relativistic system 
recovers integrability.
As in the case of the black ring,%
\footnote{The particle dynamics around a uniform circular ring 
in the 4D Euclidean space $\mathbb{E}^4$ 
is integrable~\cite{Igata:2014bga} 
while its relativistic generalization, the timelike geodesics 
in the five-dimensional (5D) black ring spacetime, 
is chaotic~\cite{Igata:2010cd,Igata:2020dow}.}
the integrable structure may be recovered in the Newtonian limit.
One of our purposes is to fill in the missing pieces
and clarifies the boundary of integrability.

To approach this problem, we focus on stable circular massive particle orbits 
in the 5D Majumdar-Papapetrou dihole spacetime, 
which are caused by the many-body effect of the sources~\cite{Igata:2020vlx,Igata:2021wwj}.
Although it should be noted that relativistic corrections in higher dimensions
can affect particle dynamics at infinity,%
\footnote{Such phenomenon is observed, for example, 
in particle dynamics on the 5D black ring 
spacetime~\cite{Igata:2020vdb,Igata:2010ye}.}
this fact suggests that stable circular orbits also exist in Newtonian gravity.
If such orbits exist in our system,
stable bound orbits inevitably appear near these orbits (discussed in detail below).
If the system is nonintegrable, then the chaos of the stable bound orbits 
can be determined by the Poincar\'e map method.

This paper is organized as follows. 
In Sec.~\ref{sec:2}, we consider the existence of stationary particle orbits 
in the Newtonian gravitational potential generated by two fixed centers 
in $\mathbb{E}^4$. 
After formulating a method for finding stationary orbits and determining their stability, 
we show the sequences of stable/unstable stationary orbits. 
In Sec.~\ref{sec:3}, we discuss the appearance of chaos for stable bound orbits 
by using the Poincar\'e map method. 
Section~\ref{sec:4} is devoted to a summary and discussions.

\section{Formulation}
\label{sec:2}
We focus on the Newtonian potential 
generated by two point mass sources fixed at different points 
in $\mathbb{E}^4$. 
Let $\bm{r}$ be a position vector. 
Let $\bm{r}=\mp \bm{a}$ denote the positions of the sources 
with mass $M_\pm$, respectively, 
where $\bm{a}$ is a nonzero constant vector.
Then the mass density distribution is 
\begin{align}
\sigma(\bm{r})=M_+ \delta(\bm{r}+\bm{a})+M_- \delta(\bm{r}-\bm{a}),
\end{align}
where $\delta$ denotes the delta function. 
Solving the Newtonian field equation 
with the source term $\sigma(\bm{r})$,%
\footnote{The following convention for the Newtonian field equation is used: 
\begin{align}
\Delta \Phi(\bm{r})=\Omega_{3} G \sigma (\bm{r}),
\end{align}
where 
$\Delta$ is the Laplacian of $\mathbb{E}^4$, and 
$\Omega_{3}=2\pi^2$ is the surface area of the unit $S^3$.}
we obtain 
a Newtonian gravitational potential
\begin{align}
\Phi(\bm{r})
=- \frac{GM_+}{2\:\!|\:\!\bm{r}+\bm{a}\:\!|^2}
- \frac{GM_-}{2\:\!|\:\!\bm{r}-\bm{a}\:\!|^2},
\end{align}
where $G$ is the gravitational constant. 
Hereafter, the masses are assumed to be equal, $M_\pm=M$.
Introduce the cylindrical coordinates $(\rho, \theta, \phi, z)$ in which 
the Euclidean metric takes the form 
\begin{align}
\mathrm{d}\ell^2
=\mathrm{d}\rho^2
+\rho^2 (\mathrm{d}\theta^2+\sin^2\theta\:\!\mathrm{d}\phi^2)
+\mathrm{d}z^2.
\end{align}
Without loss of generality, 
we may put the point sources at $z=\pm a$ on the $z$ axis, 
where $a=|\:\!\bm{a}\:\!|$. 
Then the potential $\Phi(\bm{r})$ in these coordinates is given by
\begin{align}
\Phi(\bm{r})
=-\frac{GM}{2}\left(
\frac{1}{r_+^2}+\frac{1}{r_-^2}
\right),
\end{align}
where 
\begin{align}
r_\pm=\sqrt{(z\pm a)^2+\rho^2}.
\end{align}

Let us consider freely falling particle motion in $\Phi(\bm{r})$. 
Let $m$ be particle mass and let $\bm{p}$ be a canonical momentum 
conjugate with coordinates of the particle. 
The Hamiltonian of this mechanical system is given by
\begin{align}
H
&=\frac{|\:\!\bm{p}\:\!|^2}{2m}+m\Phi
\\
&=\frac{1}{2m}\left(
p_z^2+p_\rho^2+\frac{Q^2}{\rho^2}
\right)
-\frac{\alpha\:\!m}{2}\left(
\frac{1}{r_+^2}+\frac{1}{r_-^2}
\right),
\end{align}
where $\alpha=GM$ and $Q^2$ is defined by
\begin{align}
Q^2=p_\theta^2+\frac{p_\phi^2}{\sin^2\theta},
\end{align}
which is a constant of motion associated with the $S^2$ rotational symmetry of $\Phi$.
We use units in which $m=1$ in what follows.
The Hamiltonian $H$ is equivalent to particle energy and takes a constant value $E$. 
The energy conservation $H=E$ leads to the energy equation 
\begin{align}
&\frac{1}{2} (\dot{z}^2+\dot{\rho}^2)+V=E,
\\
&V(\rho, z; Q^2)=\frac{Q^2}{2\rho^2}-\frac{\alpha}{2}\left(
\frac{1}{r_+^2}+\frac{1}{r_-^2}
\right),
\end{align}
where the dots denote the derivatives with respect to time, and 
the Hamilton equations, 
$p_z=\dot{z}$ and $p_\rho=\dot{\rho}$, have been used.
We call $V$ the effective potential in what follows.

We focus on stationary orbits where the $\rho$ and $z$ coordinates of 
a particle remain constant. 
Note that all such orbits are circular 
because of the $S^2$ rotational symmetry of $\Phi$. 
To move on the circular orbits, a particle must stay 
at a stationary point of the effective potential $V$, 
where the following conditions must hold:
\begin{align}
\label{eq:Vz}
&V_z=\alpha \left(
\frac{z+a}{r_+^4}+\frac{z-a}{r_-^4}
\right)=0, 
\\
\label{eq:Vrho}
&V_\rho=-\frac{Q^2}{\rho^3}+\alpha \rho \left(
\frac{1}{r_+^4}
+\frac{1}{r_-^4}
\right)=0,
\\
\label{eq:V=E}
&V=E,
\end{align}
where $V_i=\partial_i V$ $(i=z, \rho)$. 
The condition~\eqref{eq:Vz} leads to the 
three real roots
\begin{align}
z&=0, \\
z&=\pm z_0(\rho):=\pm \sqrt{\frac{3a^2-\rho^2}{1+2a/\sqrt{\rho^2+a^2}}},
\end{align}
and the others are imaginary roots. 
Note that $z_0$ is well-defined only in the range $0<\rho\leq \sqrt{3}a$.
Furthermore, we obtain the following values $Q_0^2$ and $E_0$ by solving 
Eqs.~\eqref{eq:Vrho} and \eqref{eq:V=E} for $Q^2$ and $E$: 
\begin{align}
&Q^2=Q^2_0:=\alpha\rho^4\left(
\frac{1}{r_+^4}
+\frac{1}{r_-^4}
\right),
\\
&E=E_0:=V(\rho, z; Q_0^2)=
\frac{\alpha \rho^2}{2}\left(
\frac{1}{r_+^4}+\frac{1}{r_-^4}
\right)-\frac{\alpha}{2}\left(\frac{1}{r_+^2}+\frac{1}{r_-^2}\right).
\end{align}
Since $Q_0^2$ is always non-negative, we can find a stationary orbit at any point 
on the sequences
\begin{align}
&\gamma_0:=\left\{\:\!(\rho, z)\setminus (0, \pm a)
\:\!\big|\:\!z=0 \ \mathrm{or} \ z=\pm z_0
\:\!\right\}.
\end{align}

Let us divide $\gamma_0$ into two parts, depending on 
whether the stationary orbit at each point on $\gamma_0$ is stable or unstable. 
If $V$ at an extremum point on $\gamma_0$ is locally minimized, 
then the circular orbit is stable. 
If not, that is, if it is locally maximized or has a saddle point, then the circular orbit is unstable.
Let $(V_{ij})$ be the Hessian matrix of $V$, 
where $V_{ij}(\rho, z; Q^2):=\partial_j \partial_i V$ ($i, j=\rho, z$). 
Let $h$ and $k$ be the determinant and the trace of $(V_{ij})$, respectively, 
i.e., $h(\rho, z; Q^2):=\det (V_{ij})$ and $k(\rho, z; Q^2):=\mathrm{tr} (V_{ij})$.
Evaluating $h$ and $k$ at $Q^2=Q_0^2$, we obtain
\begin{align}
&h_0(\rho, z)
:=h(\rho, z; Q_0^2)
=
 \frac{64 \:\!\alpha^2 a^2\rho^2}{r_+^6r_-^6}
-12\alpha^2 \left(
\frac{1}{r_+^4}+\frac{1}{r_-^4}
\right)\left[
\frac{(z+a)^2}{r_+^6}
+\frac{(z-a)^2}{r_-^6}
\right],
\\
\label{eq:k0}
&k_0(\rho, z):=k(\rho, z; Q_0^2)=\alpha \left(
\frac{1}{r_+^4}+\frac{1}{r_-^4}
\right),
\end{align}
where $V_{ij, 0}(\rho, z):=V_{ij}(\rho, z; Q_0^2)$ are given by
\begin{align}
&V_{\rho\rho,0}=
4\:\!\alpha \left(\frac{1}{r_+^4}+\frac{1}{r_-^4}\right)
-4\:\!\alpha \rho^2\left(
\frac{1}{r_+^6}+\frac{1}{r_-^6}
\right),
\\
&V_{zz,0}=
\alpha \left(\frac{1}{r_+^4}+\frac{1}{r_-^4}\right)
-4\:\!\alpha \left[\:\!
\frac{(z+a)^2}{r_+^6}+\frac{(z-a)^2}{r_-^6}
\:\!\right],
\\
&V_{\rho z, 0}=V_{z\rho,0}
=-4\:\!\alpha \rho \left(
\frac{z+a}{r_+^6}+\frac{z-a}{r_-^6}
\right). 
\end{align}
Note that $k_0>0$ everywhere. 
Hence, the circular orbits on $\gamma_0$ are 
stable if $h_0>0$, 
unstable if $h_0<0$, 
and marginally stable if $h_0=0$.

\begin{figure}[t]
\centering
\includegraphics[width=6.5cm,clip]{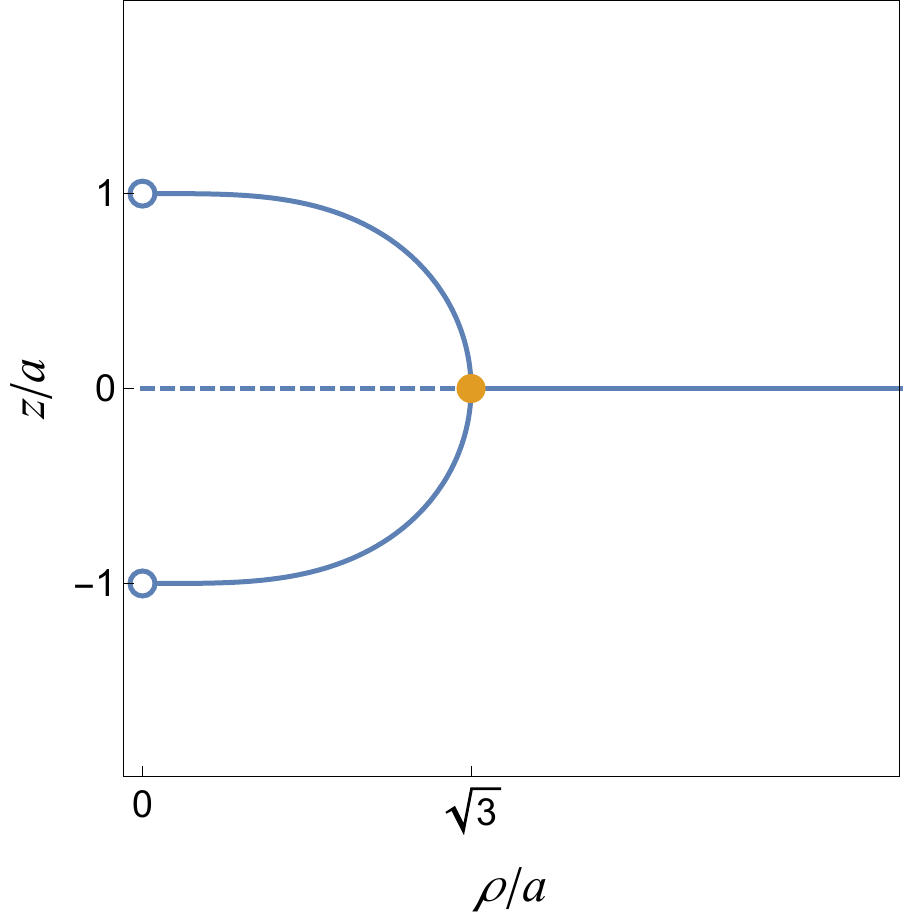}
\caption{Sequences of stable/unstable circular orbits. 
Blue solid curves show the sequences of 
stable circular orbits, and blue dashed segment shows 
the sequence of unstable circular orbits. 
Orange dot denotes the marginally stable circular orbit. 
White circles denote the locations of the point masses.}
 \label{fig:COstability}
\end{figure}

We consider the stability of the circular orbits on $z^2=z_0^2$. 
The function $h_0$ on this branch satisfies 
\begin{align}
h_0(\rho, \pm z_0)
=\alpha^2\frac{2\:\!a^3+(2\:\!a^2-\rho^2)\sqrt{\rho^2+a^2}}{16\:\!a^4\rho^4 (\rho^2+a^2)^{3/2}}\geq 0,
\end{align}
where the equality holds for $\rho=\sqrt{3}a$. 
This result indicates that all the circular orbits are stable on this branch. 
Next we consider the stability of the circular orbits on $z=0$, where 
$h_0$ reduces to 
\begin{align}
h_0(\rho, 0)=16\:\! \alpha^2a^2 \frac{\rho^2-3\:\!a^2}{(\rho^2+a^2)^6}.
\end{align}
This implies that the circular orbits are stable (i.e., $h_0>0$) for $\rho>\sqrt{3}a$, 
unstable (i.e., $h_0<0$) for $\rho<\sqrt{3}a$, 
and marginally stable (i.e., $h_0=0$) for $ \rho=\sqrt{3}a$. 
Note that this stability behavior is caused 
by switching of the stability in the $z$ direction at $\rho=\sqrt{3} a$,
\begin{align}
V_{zz, 0}(\rho, 0)
=2\:\!\alpha\frac{\rho^2-3\:\!a^2}{(\rho^2+a^2)^3}.
\end{align}
The sequence of the circular orbits and their stability are shown in Fig.~\ref{fig:COstability}. 
Blue solid curves show the sequence of stable circular orbits, 
and blue dashed segment shows the sequence of unstable circular orbits. 

It is worth noting that while there is no stable circular orbit 
for a single point mass source in $\mathbb{E}^4$, 
there are stable circular orbits for any nonzero value of $a$ 
(up to the asymptotic region).
To see the effect of $a$ on the existence of stable circular orbits, 
we consider $V$ on $z=0$,
\begin{align}
\label{eq:V0}
V(\rho, 0)=-\frac{\alpha}{\rho^2+a^2}+\frac{Q^2}{2\rho^2}.
\end{align}
If $Q^2\geq 2\alpha$, 
the potential $V(\rho, 0)$ decreases monotonically as $\rho$ increases, 
which implies that no local minimum exists. 
If 
$Q^2<2\alpha$, however, 
it always has a local minimum at $\rho=a [Q/(\sqrt{2\alpha}-Q)]^{1/2}$,
which is caused by $V\to \infty$ as $\rho\to 0$ and $V\nearrow 0$ as $\rho\to \infty$.
In the limit $Q\to \sqrt{2\alpha}$, the local minimum point goes to infinity. 
These results imply that the existence of nonzero $a$ due to the two-body effect is essential for the potential~\eqref{eq:V0} to form a local minimum point.
Therefore, the stable circular orbit can be interpreted as a result of the many-body effect.

\section{Stable bound orbits and chaos}
\label{sec:3}
We use stable stationary orbits to find 
stable bound orbits---a particle moves in a spatially bounded region 
without reaching infinity or the sources even if small perturbations are applied. 
As discussed in the previous section, 
a particle on a stable stationary orbit must stay at a local minimum point of the 
effective potential $V$. 
If the energy level $E_0$ at the local minimum point increases slightly 
(i.e., some positive energy $\Delta E>0$ is injected), 
then it will start to move away from the local minimum. 
If the energy displacement $|\Delta E|$ is small enough, 
the effective potential contour at $E_0+\Delta E$ will be closed. 
Thus, the particle with energy $E_0+\Delta E$ oscillates 
in the vicinity of the local minimum and is bounded inside the contour, 
i.e., the particle moves on a stable bound orbit. 
Therefore, the stationary stable orbits inevitably induce 
the existence of stable bound orbits in its vicinity.

\begin{figure}[t]
\centering
 \includegraphics[width=15cm,clip]{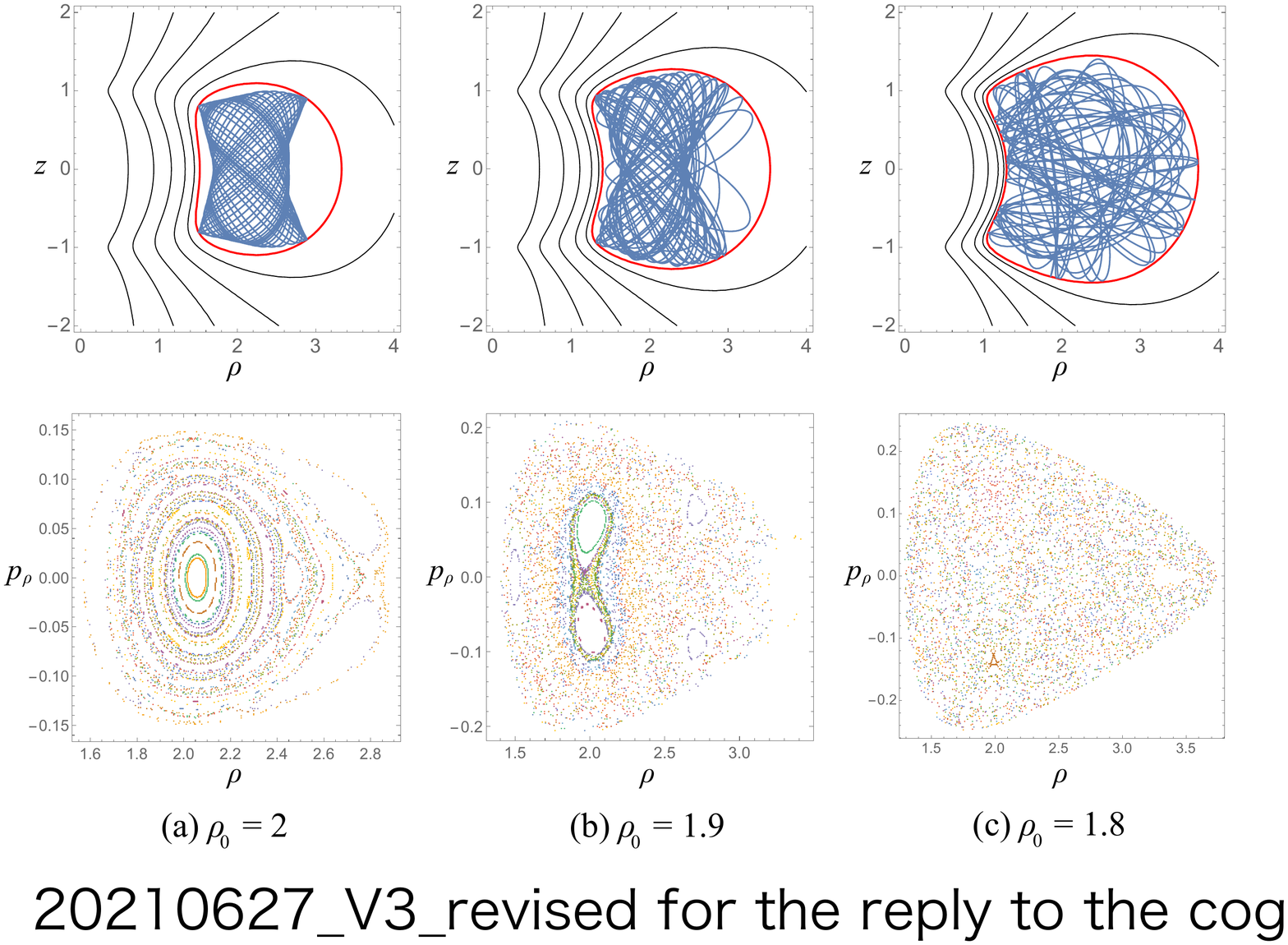}
 \caption{
Typical stable bound orbits and the Poincar\'e maps. 
Units in which $\alpha=1$ and $a=1$ are used. 
Black solid curves in the upper panels show contours of $V$ with 
$Q^2=Q_0^2(\rho_0, 0)$ in the $(\rho, z)$ plane, where 
(a) $\rho_0=2$, (b) $\rho_0=1.9$, and (c) $\rho_0=1.8$.
The position $(\rho, z)=(\rho_0, 0)$ denotes the local minimum point of $V$. 
Red solid curves correspond to the contour level with $V=-0.025$. 
Blue solid curves denote stable bound orbits 
with $E=-0.025$ and $Q^2=Q_0^2(\rho_0, z)$. 
The lower panels of the $(\rho, p_\rho)$ plane show 
the Poincar\'e maps of stable bound orbits 
for $50$ random initial conditions with the same energy and the angular momentum
as in the upper case. 
The quantities $\rho$ and $p_\rho$ are recorded 
when a particle passes through the cross section $z=0$ with $\dot{z}>0$.}
 \label{fig:chaos}
\end{figure}

Figure~\ref{fig:chaos} shows some typical effective potential contours in the upper panels. 
We use units in which $\alpha=1$ and $a=1$ in what follows. 
Black solid curves denote the contours of $V$ 
with $Q^2=Q^2_0(\rho_0, 0)$, 
where (a) $\rho_0=2.0$, (b) $\rho_0=1.9$, and (c) $\rho_0=1.8$. 
The position $(\rho,z)=(\rho_0, 0)$ corresponds to the local minimum point of $V$, 
at which the values of $E_0$ and $Q_0^2$ are evaluated as 
(a) $(E_0, Q_0^2)=(-0.04, 1.28)$, 
(b) $(E_0, Q_0^2)\simeq (-0.04705, 1.226)$,
and 
(c) $(E_0, Q_0^2)\simeq (-0.05562, 1.167)$. 
Red solid curves show the contour level with $V=-0.025$. 
Blue solid curves denote stable bound orbits with $E=-0.025$ 
and $Q^2=Q^2_0(\rho_0, 0)$, which are confined in the red closed curves. 
In the case~(a), the stable bound orbit shows ordered patterns inside the red closed curve. 
These are formed by the superposition of oscillations in two directions, a Lissajous curve. 
In the case~(b)---the local minimum point is a bit closer to the sources 
than the case (a)---the stable bound orbit shows a slightly disturbed from a Lissajous curve. 
This can be seen as evidence that the chaotic nature of the orbit begins to emerge. 
In the case (c)---the local minimum point is the closest 
among these three cases---the stable bound orbit denotes no longer a Lissajous curve. 
The completely irregular pattern suggests the emergence of chaotic nature of the orbits.

Using the Poincar\'e map method,%
\footnote{Our Hamiltonian system has two independent constants of motion, $Q^2$ and $p_\phi$, 
i.e., these are Poisson commutable with each other.
They reduce the degrees of freedom of the system from four to two. 
This fact allows to use the two-dimensional Poincar\'e map to evaluate chaos. 
Note that we should use higher-dimensional Poincar\'e map 
if the degrees of freedom of the system is larger than three~(see, e.g., Refs.~\cite{Patsis:1994,Katsanikas:2012vi}).} 
we evaluate more quantitatively the signs of chaos in stable bound orbits.
The lower panels of Fig.~\ref{fig:chaos} show the Poincar\'e maps of stable bound orbits 
for 50 different initial conditions (denoted by different colors) 
with the same energy and the angular momentum as in the upper case. 
The Poincar\'e section is defined at $z=0$, where 
$\rho$ and $p_\rho$ are recorded when a particle passes through it with $\dot{z}>0$. 
In the case~(a), we find closed dotted loops in the $(\rho, p_\rho)$ plane 
for each stable bound orbit. 
This implies that the orbits in the phase space lie on a torus, 
which means that the chaotic nature of the orbits is not manifest.
In the case~(b), the Poincar\'e sections form closed dotted curves in some cases 
while most are scattered to fill bounded regions in the $(\rho, p_\rho)$ plane.
The latter behavior is a result of the chaotic nature of stable bound orbits.
In the case (c), the point set that forms a closed dotted curve no longer appears, 
and all orbits show chaos.
Therefore, we can conclude that our generalized Euler's three-body problem 
shows chaos in general. 

\section{Summary and discussions}
\label{sec:4}
We have considered the generalized Euler's three-body problem, 
the dynamics of a freely falling particle 
in the Newtonian gravitational potential generated by point masses fixed, 
respectively, at two different points in $\mathbb{E}^4$. 
We presented the conditions under which particles remain in stationary orbits 
in terms of the effective potential 
and proposed a systematic procedure for solving them.
We used it to identify the sequences of stationary circular orbits 
in the generalized Euler problem.
Furthermore, the linear stability of each circular orbit was clarified.
As a result, we have found a family of stable circular orbits 
extending from two point masses to infinity, 
whose existence is independent of the separation between the sources.
It is worth noting that stable stationary orbits 
do not exist in the single point source case, 
i.e., the Kepler problem in $\mathbb{E}^4$~\cite{Tangherlini:1963bw}.
Therefore, we can conclude that the existence of stable circular orbits 
in this system is caused by the many-body effect of the sources.

We compare our results with the sequences of stable circular orbits 
in the 5D Majumdar--Papapetrou dihole spacetime 
with equal mass $M_*$ and separation $a_*$~\cite{Igata:2020vlx}. 
In the vicinity of each pair of sources, we can see the difference 
between Newtonian gravity, where there exist sequences of stable circular orbits 
up to an arbitrary neighborhood of the point masses, 
and general relativity, where a pair of the innermost stable circular orbits 
appear due to the relativistic effects.
On the other hand, in the asymptotic region, 
stable circular orbits exist when the dihole separation 
is large ($a_*/\sqrt{M_*}\geq \sqrt{3}$ in geometrized units), 
which is consistent with the result in our Euler problem.
However, our results do not reproduce 
the disappearance of stable circular orbits 
in the asymptotic region for small separation case ($a_*/\sqrt{M_*}<\sqrt{3}$). 
This difference is due to a higher-order multipole, 
which contains a relativistic correction in this case, 
affecting the existence of stable circular orbits in the asymptotic region 
of higher-dimensional spacetime.
Such relativistic effects on particle dynamics at infinity 
are unique to higher-dimensional spacetimes.

In the latter part of this paper, we used the stable stationary orbits 
to find stable bound orbits. 
Calculating the Poincar\'e maps for the stable bound orbits, 
we revealed the appearance of chaotic behavior. 
Furthermore, we found that the chaos increases as the region 
where the stable bound orbits lie approaches the sources.
These results imply that our generalized Euler problem is nonintegrable.
In other words, there is no additional constant of motion 
such as the Whittaker constant. 
Therefore, we conclude that the chaos observed in the timelike geodesics 
of the 5D Majumdar-Papapetrou dihole spacetimes does not disappear 
even in the Newtonian limit.

The fact that our system has no integrable structure is preserved 
even if we parameterize the spatial dimension 
without changing the inverse-square potential. 
On the other hand, if the exponent of the inverse power-law of the potential is 
larger than $2$, we need further study. 
In such cases, however, the Poincar\'e map method may not be applicable 
because no stable bound orbits are expected to exist.
Euler's three-body problem in a higher-dimensional Euclidean space with compactified 
extra-dimensional space is also an interesting problem for the future.

\begin{acknowledgments}
This work was supported by Grant-in-Aid for Early-Career Scientists 
from the Japan Society for the Promotion of Science~(JSPS KAKENHI Grant No.~JP19K14715). 
\end{acknowledgments}

\end{document}